\documentclass[aps,prl,showpacs,amsmath,amssymb,amsfonts,superscriptaddress,twocolumn,lengthcheck]{revtex4-2}
\usepackage{ulem}
\usepackage{color}
\usepackage{ulem}
\usepackage{amsmath}
\usepackage{amssymb}
\usepackage{graphics}
\usepackage{graphicx}
\usepackage{dcolumn}
\usepackage[pdfpagemode=UseNone,pdfstartview=FitH,colorlinks=true,linkcolor=blue,urlcolor=blue,anchorcolor=blue,citecolor=blue]{hyperref}
\usepackage{lineno}
\usepackage{booktabs}

\begin{document}
\title{Quantum control of a single magnon in a macroscopic spin system}

\author{Da Xu}
\thanks{These two authors contributed equally}
\affiliation{Interdisciplinary Center of Quantum Information, State Key Laboratory of Modern Optical Instrumentation and Zhejiang Province Key Laboratory of Quantum Technology and Device, School of Physics, Zhejiang University, Hangzhou 310027, China}

\author{Xu-Ke Gu}
\thanks{These two authors contributed equally}
\affiliation{Interdisciplinary Center of Quantum Information, State Key Laboratory of Modern Optical Instrumentation and Zhejiang Province Key Laboratory of Quantum Technology and Device, School of Physics, Zhejiang University, Hangzhou 310027, China}

\author{He-Kang Li}
\affiliation{Interdisciplinary Center of Quantum Information, State Key Laboratory of Modern Optical Instrumentation and Zhejiang Province Key Laboratory of Quantum Technology and Device, School of Physics, Zhejiang University, Hangzhou 310027, China}

\author{Yuan-Chao Weng}
\affiliation{Interdisciplinary Center of Quantum Information, State Key Laboratory of Modern Optical Instrumentation and Zhejiang Province Key Laboratory of Quantum Technology and Device, School of Physics, Zhejiang University, Hangzhou 310027, China}

\author{Yi-Pu Wang}
\email[Corresponding author.~Email:~]{yipuwang@zju.edu.cn}
\affiliation{Interdisciplinary Center of Quantum Information, State Key Laboratory of Modern Optical Instrumentation and Zhejiang Province Key Laboratory of Quantum Technology and Device, School of Physics, Zhejiang University, Hangzhou 310027, China}

\author{Jie Li}
\affiliation{Interdisciplinary Center of Quantum Information, State Key Laboratory of Modern Optical Instrumentation and Zhejiang Province Key Laboratory of Quantum Technology and Device, School of Physics, Zhejiang University, Hangzhou 310027, China}

\author{H. Wang}
\affiliation{Interdisciplinary Center of Quantum Information, State Key Laboratory of Modern Optical Instrumentation and Zhejiang Province Key Laboratory of Quantum Technology and Device, School of Physics, Zhejiang University, Hangzhou 310027, China}
\affiliation{Hefei National Laboratory, Hefei 230088, China}

\author{Shi-Yao Zhu}
\affiliation{Interdisciplinary Center of Quantum Information, State Key Laboratory of Modern Optical Instrumentation and Zhejiang Province Key Laboratory of Quantum Technology and Device, School of Physics, Zhejiang University, Hangzhou 310027, China}
\affiliation{Hefei National Laboratory, Hefei 230088, China}

\author{J. Q. You}
\email[Corresponding author.~Email:~]{jqyou@zju.edu.cn}
\affiliation{Interdisciplinary Center of Quantum Information, State Key Laboratory of Modern Optical Instrumentation and Zhejiang Province Key Laboratory of Quantum Technology and Device, School of Physics, Zhejiang University, Hangzhou 310027, China}

\date{\today}
\begin{abstract}
Non-classical quantum states are the pivotal features of a quantum system that differs from its classical counterpart. However, the generation and coherent control of quantum states in a macroscopic spin system remain an outstanding challenge. Here we experimentally demonstrate the quantum control of a single magnon in a macroscopic spin system (i.e., 1~mm-diameter yttrium-iron-garnet sphere) coupled to a superconducting qubit via a microwave cavity. By tuning the qubit frequency {\it in situ} via the Autler-Townes effect, we manipulate this single magnon to generate its non-classical quantum states, including the single-magnon state and the superposition of single-magnon state and vacuum (zero magnon) state. Moreover, we confirm the deterministic generation of these non-classical states by Wigner tomography. Our experiment offers the first reported deterministic generation of the non-classical quantum states in a macroscopic spin system and paves a way to explore its promising applications in quantum engineering.
\end{abstract}


\maketitle
The generation of non-classical quantum states was achieved in some macroscopic systems such as the superconducting resonator~\cite{Hofheinz-Nature-2008,Hofheinz-Nature-2009}, optomechanical resonator~\cite{Hong-Science-2017}, and acoustic-wave systems~\cite{Cleland-Nature-2018,Schoelkopf-Nature-2018}. However, it remains an outstanding challenge for a macroscopic spin system. Recently, quantum magnonics becomes a newly developed area attracting considerable interest~\cite{Lachance-APE-2019,Yuan-PhysRep-2022,Rameshti-PhysRep-2022}. It is demonstrated that ferromagnetic magnons can strongly and coherently couple to microwave photons in a cavity~\cite{Huebl-PRL-2013,Tabuchi-PRL-2014,Zhang-PRL-2014,Goryachev-PRApplied-2014,Bai-PRL-2015,Zhang-NC-2017}.
Mediated by the cavity, magnons can also couple to a superconducting qubit~\cite{Tabuchi-Science-2015}, making it realizable to resolve magnon numbers in a low-excitation coherent state of magnons~\cite{Lachance-quirion-SA-2017,Lachance-quirion-Science-2020}. These demonstrations have removed barriers towards exploring the quantum regime of a macroscopic spin system~\cite{Lachance-APE-2019}. Because the hybrid magnon-qubit system was therein operated in the {\it dispersive} regime (i.e., the magnon linewidth is comparable to the magnon-qubit dispersive interaction strength), it is still difficult to manipulate quantum states of the macroscopic spin system in this regime.

In this work, we report for the first time the quantum control of a single magnon in a macroscopic yttrium-iron-garnet (YIG) sphere. We manipulate the single magnon via a superconducting qubit that can resonantly couple to the YIG sphere in a tunable manner and {\it deterministically} generate the single-magnon state and the superposition of single-magnon state and vacuum (zero magnon) state. These states are typical quantum states of the macroscopic spin system. The coupling between the magnon and qubit is mediated by a three-dimensional (3D) microwave cavity, and the dressed Autler-Townes (AT) doublet states are used to tune the qubit frequency~\cite{AT-PR-1955,Han-PRApplied-2019}, which enables us to explore the magnon-qubit hybrid quantum system in the resonant-coupling regime. In contrast to the previous demonstrations using a dispersive interaction, this resonant coupling can give a much faster energy transfer between the qubit and the magnon, thus allowing us to implement sufficient quantum operations to generate the quantum states of a single magnon within the coherence time of the system. Our quantum control of a single magnon is precise and deterministic, making the YIG spin system one of the largest systems that become able to generate macroscopic quantum states. It paves a way to explore promising applications in quantum engineering such as the quantum transducer~\cite{Zeuthen-QST-2020,Lambert-AQT-2020,Mirhosseini-Nature-2020} and quantum network~\cite{Kimble-Nature-2008,Li-PRXQuantum-2021}.

\begin{figure}
	\centering
	\includegraphics[width=0.48\textwidth]{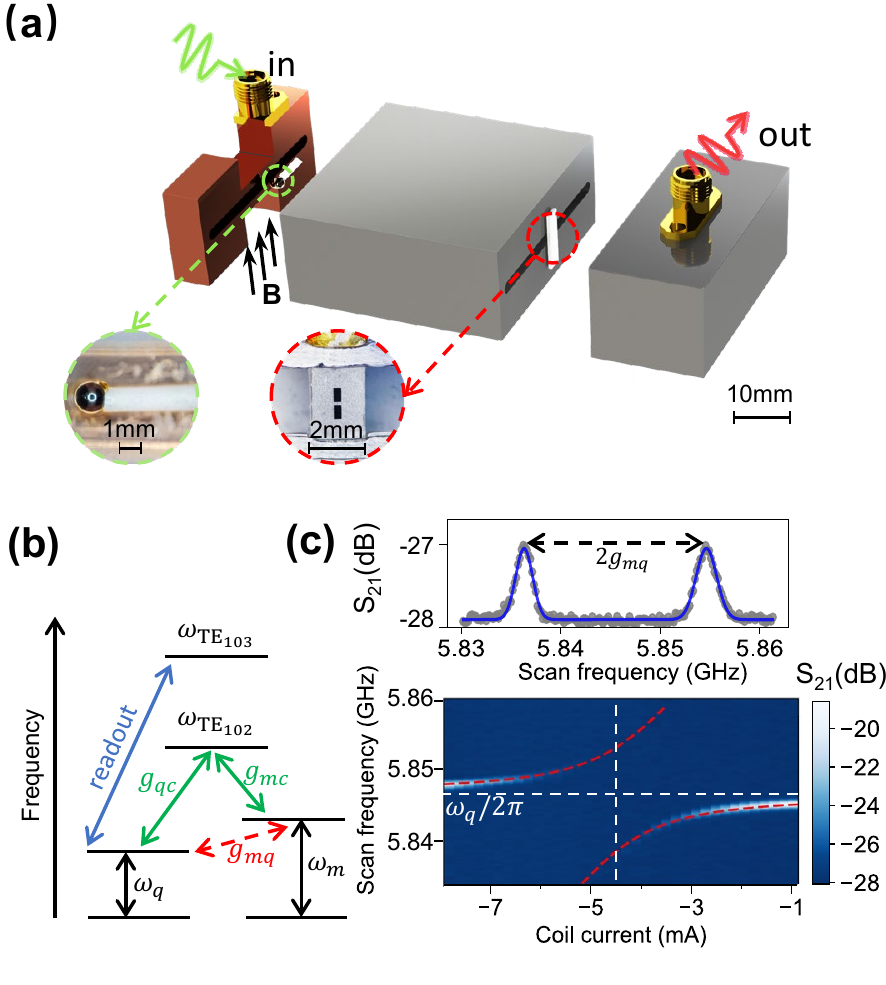}%
	\caption{(a)~A rectangular 3D cavity consists of a small part made of oxygen-free copper and a large part made of aluminium. A 1 mm-diameter YIG sphere is placed in the small part of the cavity and a 3D transmon qubit~\cite{Paik-PRL-2011,Rigetti-PRB-2012} is mounted in the large part of the cavity. The aluminium part of the cavity is further covered by an annealed pure iron magnetic shield.
The enlarged green and red dashed circles show the optical microscopy images of the YIG sphere and the qubit chip, respectively.  The YIG sphere is magnetized by an external magnetic field ${\bf B}$ (cf.~\cite{supplemental} I.A). (b)~Level structure diagram of the coupled system. Both the magnon and the qubit are directly coupled to the cavity mode $\rm{TE}_{102}$ dispersively, resulting in an effective interaction $g_{mq}$ between the magnon and the qubit.  (c) Bottom: $S_{21}$ transmission measured around $\omega_q$ by a vector network analyzer (VNA). Level anticrossing due to the interaction between the magnon and the qubit is shown, where the red dashed curves are analytical fittings (cf.~\cite{supplemental} II.A), the horizontal dashed line corresponds to the qubit frequency $\omega_q$, and the vertical dashed line corresponds to the resonant coupling point. The qubit spectrum is measured as a function of  the coil current of the electromagnet (which is proportional to the magnetic field strength). Top: $S_{21}$ transmission at the resonant point where the coil current is about  -4.5 mA. The frequency difference between two peaks is $2g_{mq}$.}
	\label{fig1}
\end{figure}

The hybrid quantum system that we study consists of a 1~mm-diameter YIG sphere and a superconducting qubit in a rectangular 3D microwave cavity [see Fig.~\ref{fig1}(a)]. The YIG sphere is placed in the copper part of the cavity and near the magnetic-field antinode of the cavity mode ${\rm TE}_{102}$, while the superconducting qubit is mounted in the aluminium part of the cavity and near the electric-field antinode of the cavity mode ${\rm TE}_{102}$.
The aluminium is superconducting at the cryogenic temperature, which can enhance both the cavity quality factor and the qubit lifetime.
The bare frequency of the cavity mode ${\rm TE}_{102}$ is about $\omega_{{\rm TE}_{102}}/2\pi=6.388$~GHz and the qubit used is a  superconducting transmon, i.e., a capacitively shunted Josephson junction, which has transition frequency $\omega_q/2\pi=5.846$~GHz and anharmonicity  $\eta/2\pi=-0.354$~GHz (see ~\cite{supplemental} I.A).

\begin{figure}
	\includegraphics[width=0.48\textwidth]{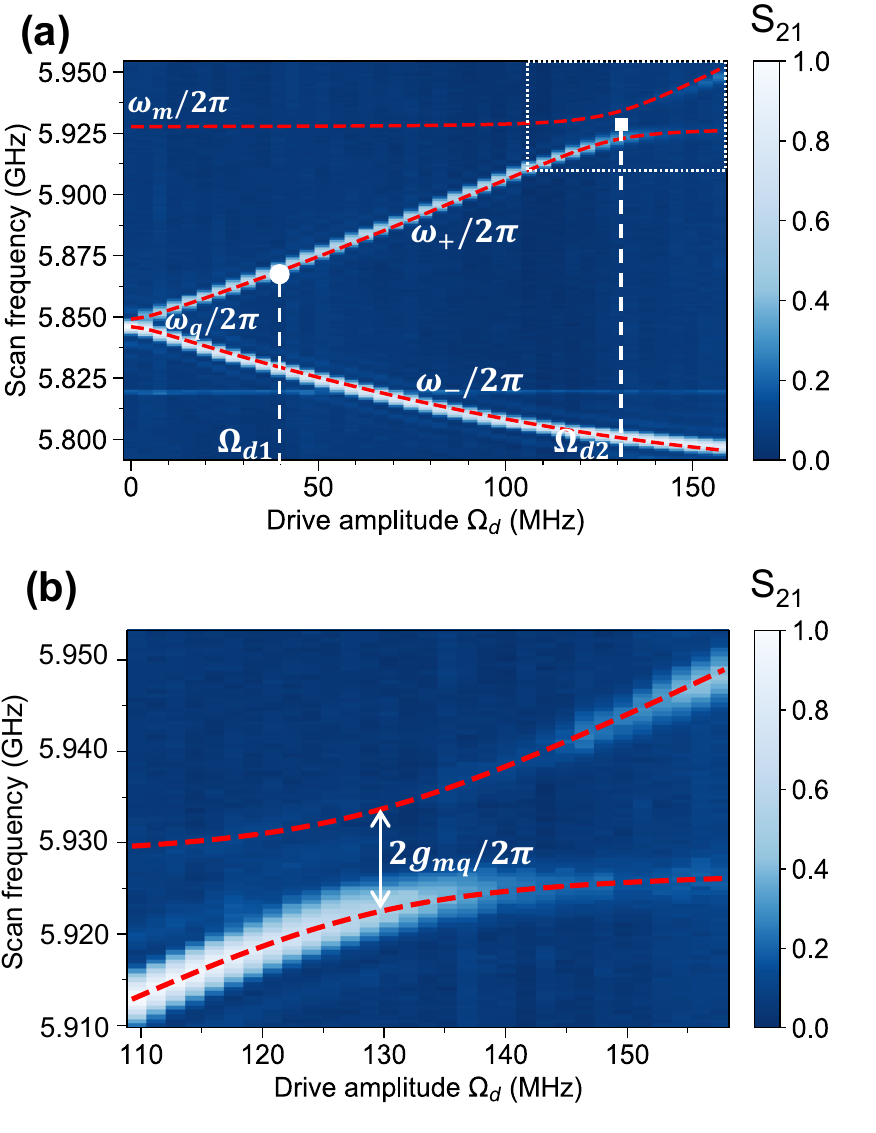}%
	\caption{(a)~$S_{21}$ measurement as a function of the drive amplitude $\Omega_d$.
A strong control drive near resonance with $\omega_{ef}$ yields the Autler-Townes splitting. The two branches denoted as $\omega_\pm/2\pi$ correspond to the transitions from ground state $\vert g,N\rangle$ to the two new eigenstates $\vert \pm,N\rangle$,
where the red dashed curves are the numerical fittings (cf. \cite{supplemental} II.A and II.B). In the experiment, the upper branch is used to tune qubit frequency.
The white dotted box at the right upper corner shows the avoided crossing between the magnon and the qubit. The white dot at the drive amplitude $\Omega_{d1}=40$~MHz is the ``work point" to create qubit excitation and the corresponding frequency is $5.870$~GHz. The white square at the drive amplitude $\Omega_{d2}=131$~MHz is the ``swap point" to implement qubit-magnon resonant swapping and the corresponding frequency is $\omega_m=5.928\ \rm{GHz}$.
(b)~Enlarged figure of the white dotted box in (a), which shows the avoided crossing corresponding to the coherent interaction between the magnon and the qubit. The coupling strength is fitted to be $g_{mq}/2\pi=5.55$~MHz.}
	\label{fig2}
\end{figure}

Direct interaction between the Kittel mode and the qubit is negligible, but strong interaction between them can be mediated by the cavity mode $\rm{TE_{102}}$ via the exchange of virtual photons~\cite{Imamoglu-PRL-2009}. In Fig.~\ref{fig1}(c), $S_{21}$ transmission is shown around $\omega_q$. Since $\omega_q$ is far detuned from $\omega_{{\rm TE}_{103}}$, the readout is performed using the Jaynes-Cummings nonlinearity readout scheme~\cite{Lachance-quirion-Science-2020,Reed-PRL-2010} via the TE$_{103}$ mode, as described in \cite{supplemental} III.A. The contour plot gives the coherent coupling strength $g_{mq}/2\pi=5.55$~MHz, as shown by the cut at constant coil current -4.3~mA. At the idle point of the qubit (i.e., without applying the control field for AT splitting), we need to suppress the magnon-qubit interaction, so we tune the Kittel mode away from the qubit in frequency and fix it at $\omega_m/2\pi=5.928$~GHz, except for the scan shown in Fig.~\ref{fig1}(c). At the idle point, the qubit lifetime and pure dephasing time are measured to be $T_{1,q}=3.65\pm0.02$~$\mu$s and $T_{\phi}=9.20\pm0.18$~$\mu$s, respectively (cf.~\cite{supplemental} III.B).
The lifetime of the Kittel mode is $T_{1,m}=128\pm2$~ns, corresponding to the magnon linewidth of $\gamma_m=1.24$~MHz (cf.~\cite{supplemental} III.D).

In order to generate the quantum states of the magnon, we should operate the magnon-qubit hybrid system in the resonant-coupling regime $\omega_q\approx\omega_m$. Conventionally, one can achieve this by making the qubit frequency tunable using a SQUID to replace the single Josephson junction in the transmon, but the quantum coherence of the qubit can be much reduced by the strong noise due to the bias magnetic field applied to the YIG sphere.
Therefore, we introduce a new technique by using the second excited state $|f\rangle$ of the transmon. With this second excited state included, the transmon becomes a three-level system, i.e., a qutrit. When the $\vert e\rangle$ to $\vert f\rangle$ transition is driven by a strong control field $\omega_{d}$, the original transition from $\vert g\rangle$ to $\vert e\rangle$ splits. This is known as the AT splitting~\cite{AT-PR-1955,Han-PRApplied-2019}. The two new eigenstates under the drive field (i.e., the AT doublet states) are $\vert +,N\rangle=\cos\theta\vert e,N\rangle+\sin\theta\vert f,N-1\rangle$ and
$\vert -,N\rangle=\sin\theta\vert e,N\rangle-\cos\theta\vert f,N-1\rangle$, with $\tan\theta=\Omega_d/(\sqrt{\Delta_d^2+\Omega_d^2}-\Delta_d)$, where $N$ is the photon number of the control field, $\Delta_d\equiv\omega_{ef}-\omega_d$ is the frequency detuning between the control field and the $\vert e\rangle$ to $\vert f\rangle$ transition, and $\Omega_d$ is the Rabi frequency related to the control-field amplitude (cf. ~\cite{supplemental} II.B). The transition frequencies from $\vert g,N\rangle$ to $\vert \pm,N\rangle$ are $\omega_{\pm}=\omega_q+\Delta_d/2\pm\sqrt{\Delta_d^2+\Omega_d^2}/2$.
Our qutrit works under the near-resonance condition $\Omega_d\gg \vert\Delta_d\vert$, where the two new eigenstates are closely reduced to $\vert \pm,N\rangle=(\vert e,N\rangle\pm\vert f,N\!-\!1\rangle)/\sqrt{2}$, i.e., nearly irrelevant to $\Omega_d$.
We define $\vert g,N\rangle$ and $\vert+,N\rangle$ as the two basis states of the qubit under the control field. Therefore, we achieve a frequency-tunable qubit by tuning the control-field amplitude.

\begin{figure}
	\includegraphics[width=0.48\textwidth]{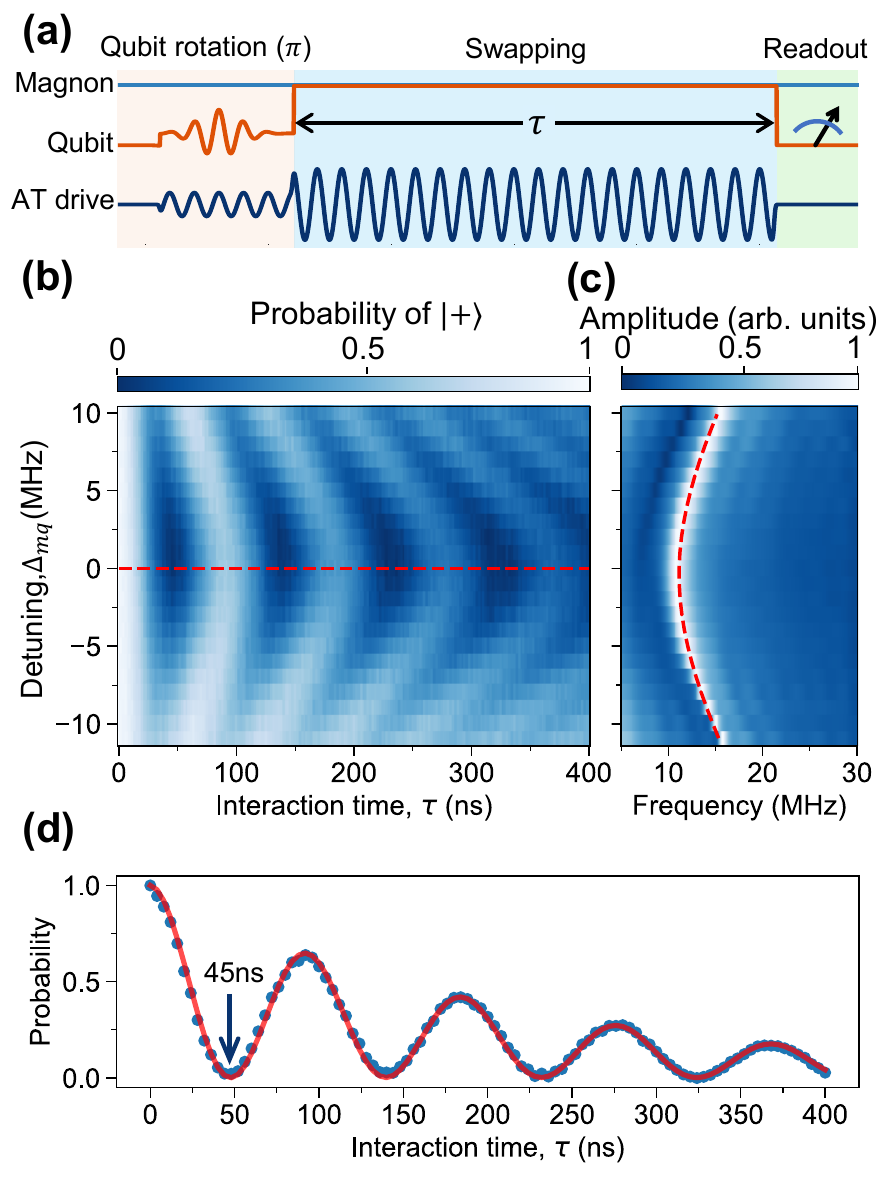}%
	\caption{(a)~Operation sequence for the qubit-magnon swapping. The straight lines refer to the frequency values of the qubit, magnon and AT drive, while perturbations sketch applied microwave pulses. Both magnon and qubit are initialized in their ground states $\vert 0\rangle$ and $\vert g\rangle$, respectively. Subsequently, the qubit is tuned to the ``work point" with transition frequency $\omega_r=5.870$~GHz using the drive amplitude $\Omega_{d1}$, cf.~Fig.~\ref{fig2}(a). A $\pi$ pulse rotates the qubit to the excited state $\vert +,N\rangle$, and then we use the drive amplitude $\Omega_{d2}$, cf.~Fig.~\ref{fig2}(a), to tune the qubit in resonance with the Kittel mode for qubit-magnon swapping. Readout of the qubit states is finally implemented at the idle point of the qubit. (b)~Qubit excited-state probability $P_+$ versus the interaction time $\tau$. The Chevron pattern shows the qubit-magnon swapping. (c)~Fourier transform of the Chevron pattern in (b), where the red dashed curve is the fitting result using $\sqrt{4g_{mq}^2+\Delta_{mq}^2}$, with $g_{mq}/2\pi=5.55\rm{MHz}$. (d)~Qubit-magnon swap curve in the resonant case, corresponding to the red dashed line in (b).
}
	\label{fig3}
\end{figure}

Our first step is to characterize the AT splitting for tuning the qubit frequency. The control field is set at $\omega_d/2\pi=5.489$~GHz,  which is near resonance with
the $\vert e\rangle \rightarrow\vert f\rangle$ transition frequency $\omega_{ef}/2\pi=5.492~\rm{GHz}$. We choose a negative detuning ($\omega_d<\omega_{ef}$) to have $\omega_d$ more off-resonant with the qubit transition frequency $\omega_q$ to reduce the unwanted excitation of $\vert g\rangle$ by the strong control field. As shown in Fig.~\ref{fig2}(a), the frequency difference between the two AT doublets $\vert\pm,N\rangle$ increases with the amplitude of the control field $\Omega_d$ (see ~\cite{supplemental} II.B on how $\Omega_d$ is obtained). Next, we measure the magnon-qubit coherent interaction in both frequency and time domains. The avoided crossing in Fig.~\ref{fig2}(b) is the magnon-qubit vacuum Rabi splitting. It is an alternative version of Fig.~\ref{fig1}(c). The qubit-magnon swapping measurement in the time domain is also performed [see Figs.~\ref{fig3}(a) and \ref{fig3}(b)]. The qubit is excited at the ``work point" with transition frequency $\omega_r=5.870$~GHz, and the corresponding control-field amplitude is $\Omega_{d1}=40$ MHz. The Chevron pattern shows the coherent exchange between the qubit and magnon states, where the oscillation frequency is fitted to $\sqrt{4g_{mq}^2+\Delta_{mq}^2}$, with $g_{mq}/2\pi=5.55$~MHz [see Fig.~\ref{fig3}(c)]. The swap curve at the resonant condition is given in Fig.~\ref{fig3}(d), which shows that a full qubit-magnon swapping takes 45~ns. Such a swapping takes a time much shorter than the magnon lifetime $T_{1,m}= 128\pm2$~ns. This is a prerequisite for high-fidelity magnon state generation and benchmarking.

\begin{figure*}
\includegraphics[width=0.85\textwidth]{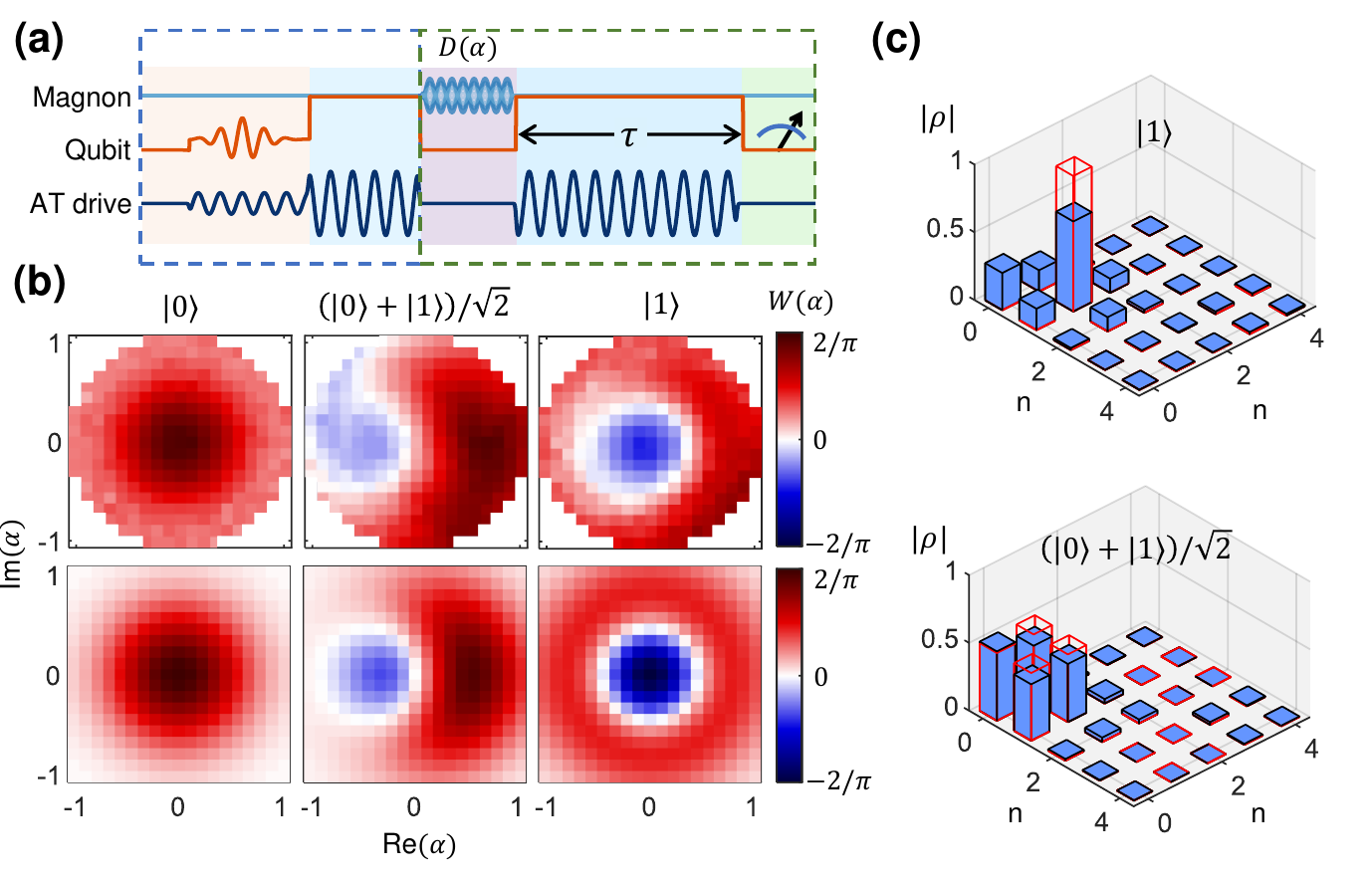}%
\caption{(a)~Operation sequences for magnon-state generation and Wigner tomography, which are shown in the blue and green dashed boxes, respectively. Both magnon  and qubit are initialized in their ground states. Subsequently, the qubit is tuned to the work point using a small-amplitude control drive. A pulse with certain amplitude and phase (e.g., the $\pi/2$ or $\pi$ pulse) is applied to rotate the qubit to the required state, and then a larger-amplitude control drive is used to have the qubit in resonant interaction with the Kittel mode for 45~ns to implement the magnon-qubit swapping. Immediately, a pulse with given amplitude and phase is applied to the Kittel mode to achieve a displacement operator $D(\alpha)$ on the magnon. Right after this, the qubit is tuned in resonance with the Kittel mode again for a period of time $\tau$ using a larger-amplitude control drive. Readout of the qubit states is finally implemented at the idle point. The swap curves, as obtained in Fig.~\ref{fig3}, are fitted using multiple linear regression to give the Wigner function.
(b)~Top row:~Wigner tomography data for three magnon states $\vert 0\rangle$, $(\vert 0\rangle+\vert 1\rangle)/\sqrt{2}$, and $\vert 1\rangle$.
Bottom row: Analytic results of the corresponding ideal magnon states. (c)~Reconstructed density matrices for the magnon states $\vert 1\rangle$ and $(\vert 0\rangle+\vert 1\rangle)/\sqrt{2}$.
The red frame bars are analytical results of the corresponding ideal magnon states. The number of shots is $8.25\times10^4$ for this measurement.}
\label{fig4}
\end{figure*}

We now generate non-classical magnon states. The operation sequence is shown in Fig.~\ref{fig4}(a). Initially, both the magnon and qubit are prepared in the ground state $\vert 0\rangle\otimes\vert g\rangle$. Then we generate the AT doublet states with drive amplitude $\Omega_{d1}$ and use a $\pi$ pulse to excite the qubit from $\vert g,N\rangle$ to $\vert +,N\rangle$. Afterwards, we tune the qubit frequency to the swap point with drive amplitude $\Omega_{d2}$, cf. Fig.~\ref{fig2}(a), to have the qubit in resonance with the magnon mode for a full magnon-qubit swapping. Then, the magnon is prepared into the single-magnon state $|1\rangle$. Similarly, we can generate a superposition state of the single magnon and vacuum, $(\vert 0\rangle+\vert 1\rangle)/\sqrt{2}$, as follows. First, at the ``work point",
we prepare the qubit in the superposition state $(\vert g,N\rangle +\vert +,N\rangle)/\sqrt{2}$ by applying a $\pi/2$ pulse to the qubit. Then, we tune the control-field amplitude to resonantly couple the qubit to the magnon mode for 45~ns.
We can also prepare the qubit in an arbitrary superposition state $\frac{1}{\mathcal{N}}(\vert g,N\rangle+c\vert +,N\rangle)$ by using a pulse with certain amplitude and phase, where $c$ is a complex number and $\mathcal{N}=\sqrt{1+|c|^2}$. Then, swapping the qubit state into the magnon, we achieve an arbitrary single-magnon superposition state $\frac{1}{\mathcal{N}}(\vert 0\rangle+c\vert 1\rangle)$. Here, we take the equal-amplitude superposition state $(\vert 0\rangle+\vert 1\rangle)/\sqrt{2}$ as a typical example.
At the work and swap points, $\Omega_d$ is found to be 40~MHz and 131~MHz, respectively, while $\Delta_d=3$~MHz. The near-resonance condition $\Omega_d\gg \Delta_d$ is satisfied in the region between these two points, where the qubit is manipulated to generate the magnon states.

Finally, we perform the Wigner tomography to characterize the generated magnon states. The Wigner function is expressed as $W(\alpha)=(2/\pi) {\rm Tr}[D(-\alpha)\rho D(\alpha)P]$, where $\rho$ is the density matrix of the generated magnon state, $D(\alpha)$ is the magnon displacement operator, and $P=e^{i\pi b^\dagger b}$ is the magnon parity operator, with $b$ ($b^\dagger$) being the annihilation (creation) operator of the magnon. Experimentally, we obtain the Wigner function by measuring parities of the displaced magnon states. The displacement operator $D(\alpha)$ is applied using a drive at $\omega_d=\omega_m$. After the magnon displacement operation, the magnon is in a superposition of many Fock states. We bring the qubit into resonance with the magnon for a period of time $\tau$ [see Fig.~\ref{fig4}(a)], having the qubit interact with all the occupied Fock states of the magnon. The qubit excited state $\vert+,N\rangle$ is subsequently read out. The resulting swap curve is the probability of the qubit excited state after the period $\tau$ of the interaction time. Fitted to the experimental data with numerically simulated swap curves, we can obtain the diagonal density matrix elements of the displaced magnon states and then use them to evaluate the Wigner function (cf.~\cite{supplemental} III.F). The Wigner tomography for three generated magnon states are shown in Fig.~\ref{fig4}(b). We can reconstruct the magnon-state density matrix from the measured Wigner functions; see Fig.~\ref{fig4}(c) and \cite{supplemental} III.H. Then, we obtain the fidelity of the generated magnon state via $F=\sqrt{\langle \psi\vert \rho\vert\psi\rangle}$ using the reconstructed density matrix $\rho$, with respect to the ideal magnon states $\vert \psi\rangle\equiv\vert0\rangle$, $\vert1\rangle$ and $(\vert0\rangle+\vert1\rangle)/\sqrt{2}$. The reconstructed density matrices give the state fidelities  $0.977\pm0.003$, $0.815\pm0.008$ and $0.942\pm0.009$ for the magnon vacuum state $\vert 0\rangle$, single-magnon state $\vert 1\rangle$, and superposition state $(\vert 0\rangle +\vert 1\rangle)/\sqrt{2}$, respectively. Here, to improve the generation fidelity of the superposition state $(\vert0\rangle+\vert1\rangle)/\sqrt{2}$, we compensate the energy loss during the first swapping process by exciting the qubit with a slightly larger amplitude pulse.

In conclusion, we have deterministically generated and benchmarked the non-classical quantum states of the magnon, including the single-magnon state and the equal-amplitude superposition of single-magnon state and vacuum (zero magnon) state.
With either an enhanced  magnon lifetime or magnon-qubit coupling strength, the extension of our protocol to generating arbitrary quantum states of more magnons is within the reach in the near future. In fact, the magnon linewidth can be improved with a higher-quality YIG sphere and the coupling strength can be increased by harnessing a smaller microwave cavity. Our experiment provides the possibility of utilizing quantum states of the magnon in a ferrimagnetic YIG system to implement quantum information processing~\cite{Blanter-PRL-2022,Sharma-PRB-21}, because it can be used to couple quantum systems in a diverse range of frequency, such as the microwave photons~\cite{Huebl-PRL-2013,Tabuchi-PRL-2014,Zhang-PRL-2014}, optical photons~\cite{Hisatomi-PRB-2016,Osada-PRL-2016,Zhang-PRL-2016, Haigh-PRL-2016}, and phonons~\cite{Zhang-SA-2016,Potts-PRX-2021,Shen-PRL-2022}.
Combined with the photon conversion in the YIG sphere, it is also promising to build a quantum transducer that transmits quantum information from a qubit in the microwave regime to the optical photon in a quantum network.

\begin{acknowledgments}
This work is supported by the National Key Research and Development Program of China (Grant No.~2022YFA1405200), the National Natural Science Foundation of China (Grants Nos.~92265202, 11934010, 12174329), Zhejiang Province Program for Science and Technology (Grant No.~2020C01019), the Fundamental Research Funds for the Central Universities (No.~2021FZZX001-02) and the China Postdoctoral Science Foundation (Grant No.~2019M660137).
\end{acknowledgments}

\end{document}